\documentstyle[12pt]{article}
\begin{document}
\begin{center}

{\bf Self-interaction effects on screening in three-dimensional QED}\\
\vskip 2cm
Subir Ghosh
\footnote {Email address: <subir@boson.bose.res.in>}\\
Physics Department,\\
Dinabandhu Andrews College, Calcutta 700084,\\
India.\\
\end{center}
\vskip 2cm
{\bf Abstract:}\\
We have shown that self interaction effects in massive quantum
electrodynamics can lead to the formation of bound states of quark
antiquark pairs. A current-current fermion coupling term is introduced,
which induces a well in the potential energy profile. Explicit expressions of
the effective potential and renormalized parameters are provided.
\vskip .5cm

\newpage

In a recent Letter \cite {ab} Abdalla and Banerjee have discussed the
confinement and screening problems in three dimensional QED. They have
studied the inter-"quark" potential between two static test charges in
a theory of dynamical fermions of mass $m$ coupled to electromagnetism.
Their results indicate that for small separation, the quantum potential
tends to the classical logarithmic Coulomb potential. However, for
large distance the potential tends to zero. This shows quite conclusively
the confining and screening nature of the potential. These results also
corroborate the two dimensional results \cite { r} nicely.

The present Letter is aimed at studying the stability of the above scenario,
in three dimensions,
in the presence of self-interaction among the fermions. In particular, we
have chosen the well studied current current Thirring interaction. When
the Thirring coupling $g$ is {\it positive}, the new model shows a marked
departure of a qualitative nature from \cite {ab} in the short distance
regime. In the potential profile, there appears more structure, in the
form of a {\it well}, indicating a strong repulsion below some critical distance.
This might lead to stable bound states of the quark antiquark pair.
The large distance behaviour shows the expected screening. For negative $g$,
nothing of the above dramatic nature occures, albeit the potential decreases
more sharply for short distance.

We formulate the problem along the lines of \cite {ab}. The fermion modes
in the gauged Thirring model are integrated out to incorporate quantum
(fermion loop)
effects in the subsequent classical analysis. This bosonization is done
in the large $m$ approximation. The auxiliary field $B_\mu $, introduced
to linearize the Thirring term,
is next integrated, resulting in a generalized Maxwell Chern Simons gauge
theory \cite{k} \cite{s}. The theory now contains two independent
parameters, $m$ and $g$.
 In order to gain further insight, we expand the results in powers
of $g$ and keep terms up to $O(1/m,~g,~g^2,~g/m)$. Surprisingly, the
terms linear in $g$ does not alter the results very much whereas
effect of the higher order
corrections is substantial, as mentioned earlier. This is our main result.

The ideas of screening and confinement play a central role in gauge theory.
The computational hurdles in four dimensions compel us to study the lower
dimensional models. But one has to extract results which are not artefacts
of low dimensionality and can be carried on to the real world. Previously,
in two dimensional QED, \cite {r} obtained results indicating screening
and confinement for massless and massive fermions respectively. QCD
was studied by \cite {gr}, where apart from the dynamical fermion mass, the
representations of the dynamical fermions and test charges
became important. The
problems regarding $\theta$-vacuum, screening, confinement and chiral
condensate in two and three dimensional QCD were discussed in \cite {g}.

The parent model is
\begin{equation}
L_F=\bar\psi i\gamma^\mu (\partial_\mu-ieA_\mu)\psi -m\bar\psi\psi
+{g\over 2}\mid\bar\psi\gamma^\mu\psi\mid^2 -{{pe^2}\over 4}
\mid A_{\mu\nu}\mid^2
+{{qe^2}\over 2}\epsilon_{\mu\nu\lambda}A^\mu A^{\nu\lambda}
+J_\mu A^\mu.
\label{eqlf}
\end{equation}
Here $A_{\mu\nu}=\partial_\mu A_\nu-\partial_\nu A_\mu $, $J_\mu $ is
an external conserved current and conventionally one takes $p=1/e^2$,
$q=\mu/(2e^2)$. We have considered them arbitray to keep track of them.
The above model is rewritten with the auxiliary field $B_\mu$ as
\begin{equation}
L_F=\bar\psi i\gamma^\mu (\partial_\mu-ieA_\mu -iB_\mu)\psi
-{1\over{2g}}\mid B_\mu \mid^2 -m\bar\psi\psi
-{{pe^2}\over 4}
\mid A_{\mu\nu}\mid^2 +{{qe^2}\over 2}\epsilon_{\mu\nu\lambda}A^\mu A^{\nu\lambda}
+J_\mu A^\mu.
\label{eqlfb}
\end{equation}
The bosonized lagrangian to $O(1/m)$ is
$$
L_B=-{a\over 4}\mid B_{\mu\nu}\mid^2
+{{\alpha}\over 2}
\epsilon_{\mu\nu\lambda}B^\mu B^{\nu\lambda}-{1\over{2g}}
\mid B_\mu \mid^2
-{{(a+p)}\over 4}e^2 \mid A_{\mu\nu}\mid^2
$$
\begin{equation}
+{{(\alpha+q)}\over 2}e^2
\epsilon_{\mu\nu\lambda}A^\mu A^{\nu\lambda}-{{ae}\over 2}A_{\mu\nu}B^{\mu\nu}
+e\alpha\epsilon_{\mu\nu\lambda}B^\mu A^{\nu\lambda} +J_\mu A^\mu,
\label{eqlb}
\end{equation}
where $\alpha =1/(8\pi)$ and $a=-1/(6\pi m)$. The above Lagrangian is
quadratic in $B_\mu $ and
after a formal integration of it, we get
the gauge invariant effective action,

$$
Z(A_\mu)=\int{\cal D}\delta (\partial_\mu A^\mu)exp(-i/4)[2e^2A_\mu
{{4\alpha^2g-a+a^2g\partial^2}\over{(ag\partial^2-1)^2+
4\alpha^2g^2\partial^2}}(\partial^2g^{\mu\nu}-\partial^\mu\partial^\nu)
A_\nu$$
\begin{equation}
+e^2A_\mu {{\{-4\alpha+8g(1-\alpha )
\partial^2(2\alpha^2g-a+a^2g\partial^2)\}}\over{(ag\partial^2-1)^2+
4\alpha^2g^2\partial^2}}\epsilon^{\mu\nu\lambda}\partial_\nu A_\lambda
 -4J_\mu A^\mu].
\label{eqzda}
\end{equation}
Lorentz gauge is adopted in
defining (\ref{eqlb}). This is a generalized Maxwell Chern Simons type
of theory \cite {d, s, k}.
For $g=q=0$, this action reduces to the one in \cite {ab}. From the
$A_\mu $ equation of motion, it follows that $\partial_\mu A^\mu=0$.
Let us from now on work with the truncated
version of this model keeping only terms of
$O(a,~g,~g^2,~ag~)$. The $A_\mu$-equation in Lorentz gauge is
\begin{equation}
P\epsilon_{\mu\nu\lambda}\partial^\nu A^\lambda +Q\partial^2A_\mu +J_\mu
=0,
\label{eqa}
\end{equation}
$$P=2e^2[(\alpha +q)-8(2\alpha^2g^2 -ag)(1+q)\partial^2];~~
Q=-e^2[(4\alpha^2g-a-p)+2p(2\alpha^2g^2-ag)\partial^2].
$$
The above equation can be rewritten as \cite {ab}
\begin{equation}
(\partial^2+({P\over Q})^2)(-\epsilon_{\mu\nu\lambda}\partial^\nu A^\lambda )
={P\over{Q^2}}J_\mu +{1\over Q}\epsilon_{\mu\nu\lambda}\partial^\nu J^\lambda .
\label{eqaa}
\end{equation}
For $\mu =2$ in the static case, (\ref{eqaa}) reduces to
\begin{equation}
(\partial^2+({P\over Q})^2)A_0={1\over Q}J_0.
\label{eqa0}
\end{equation}
This equation is inverted to get $A_0$.
We now proceed exactly as in \cite {ab}. The potential energy between
two external static charges, (emerging from $J_0$), a distance $L$ apart, is
\begin{equation}
V(L)=-(L_q-L_0)=-q[A_0(x^1=-L/2,~x^2=0)-A_0(x^1=L/2,~x^2=0)].
\label{eqa00}
\end{equation}
The subscript $q$ denotes the presence of the external charge $q$.
Solving the $A_\mu$-equation for $A_0$ we get
$$
A_0(x)=q_1[\Delta (x^1+L/2,x^2;N)-\Delta(x^1-L/2,x^2;N)]$$
\begin{equation}
+q_2\partial_i\partial_i[\Delta (x^1+L/2,x^2;N)-\Delta(x^1-L/2,x^2;N)].
\label{eq0}
\end{equation}
$\Delta(x,m)$ is the two dimensional Euclidean Feynman propagator
given by the modified Bessel function
$$
\Delta(x,m)=1/(2\pi)K_0(m\sqrt{(x^1)^2+(x^2)^2}).
$$
Here $q_1$ and $q_2$ and $N$ are related to the external charge $q$ and
the fermion mass $m$. Once again for $g=q=0$ we get back the \
results of \cite {ab}. The interesting term is the latter one
in (\ref{eqa0}) which disappears if
{\it only} $O(a,~g)$ terms are kept. Because of the derivative operator,
it radically alters the functional form of $A_0$, even though its strength
$ q_2$ is much less
than $q_1$. For small
separation $L$, $K_0(L)\approx ln(L) $ and the second term $ \approx
1/L^2 $, which competes and finally dominates over the first term.
This is one of our main results. Finally the potential
energy function is expressed in terms of dimensionless variables
$M\equiv e^2/m,~G\equiv ge^2,~\theta\equiv \mu/e^2,~ X=e^2L $ as
$$
V_{Th}=-{{q^2}\over\pi}{{(1+t)}\over{1-M/(6\pi)}}K_0[{{(1+u)X}\over
{4\pi(1-M/(6\pi))}}]$$
\begin{equation}
 +{{q^2}\over \pi}{{2[G^2/(32\pi^2)+GM/(6\pi)]}
\over{1-M/(3\pi)}}(\partial_X)^2K_0[{{(1+u)X}\over
{4\pi(1-M/(6\pi))}}].
\label{eqv}
\end{equation}
This expression is to be compared with $V_{classical}$ and $V_{AB}$
in \cite{ab}
$$V_{Cl}={{q^2}\over \pi}ln(X),$$
$$V_{AB}=-{{q^2}\over\pi}{1\over{1-M/(6\pi)}}K_0[{X\over
{4\pi(1-M/(6\pi))}}].$$
The correction terms  in (\ref{eqv}) are
$$u={4\over{1-M/(2\pi)}}[{G\over{64\pi^2}}+G^2(-{1\over{32\pi^3}}
-{\theta\over{128\pi^3}}+{\theta\over{8\pi^2}}$$
$$-{{3\theta^2}\over{512\pi^3}})
-MG({1\over{96\pi^3}}-{1\over{6\pi^2}}+{\theta\over{24\pi^2}}-{{2\theta}\over
{3\pi}}-{{\theta^2}\over{4\pi}})]$$
$$t={4\over{1-M/(2\pi)}}[{G\over{64\pi^2}}+G^2(-{1\over{1024\pi^4}}-{3\over
{512\pi^3}}-{\theta\over{64\pi^3}}+{\theta\over{4\pi^2}}+{{3\theta^2}
\over{32\pi^2}})$$
$$-GM({1\over{64\pi^3}}+{1\over{3\pi^2}}+{\theta\over{12\pi^2}}
-{{4\theta}\over{3\pi}}-{{\theta^2}\over{2\pi}})].$$
This constitues our main result.

Notice that $\theta $ does not play any significant
role in the present context and so we put $\theta =0$. In Figure (1), for
$M=0.1,~G=\pm0.1$, $V_{Cl},~~V_{AB}$ and $V_{Th}^\pm $ are plotted.
For this case, the potential well in $VThP(G=+0.1)$ is at $X\approx
2{\sqrt{G^2/(32\pi^2)+GM/(6\pi)}}\approx 0.046 $ and for  $X<0.003$,
$VThP$ becomes positive.
$VThN(G=-0.1)$ shows a sharper descent at short distance.
Figure (2) shows only $O(G)$ corrections
to the $V_{AB}$ result. The fact that $V_{Th}\approx V_{AB}$ up to $O(G)$
is quite insensitive for a wide range of values of $G$ and $M$. Here,
an unrealistic value $G=\pm 50$ is chosen just to separate the $V_{Th}$
and $V_{AB}$ lines as well as to stress the stability of this formulation
up to $O(M,~G)$. The shift of $V_{Th}$ with respect to $V_{AB}$ in the
upward or downward direction direction is dictated by the sign of $G$.

Up to $O(M,~G)$ the effective mass of the gauge particle $A_\mu $ and
the renormalized charge $q^2$ are
$$M_{Th}={P\over Q}\approx {{e^2/(4\pi)}\over{1-M/(6\pi)}}
(1-{G\over{2\pi}})=M_{AB}(1-{G\over{2\pi}}),$$
$$(q^2)_{Th}\approx
{{q^2}\over{1-M/(6\pi)}}
(1+{G\over{16\pi^2}})=(q^2)_{AB}(1+{G\over{16\pi^2}}).$$
Higher order momentum dependent terms have been left out.

Let us summarise our results. The potential energy between
the external charges at large
distance is screened as the gauge particles aquire mass from the Chern Simons
term. Without the Thirring interaction,
at short distance, one gets a decreasing negative
potential, logarithmic in nature. The
positive Thirring coupling term introduces a sort of
centrifugal barrier in the effective potential, which leads to the potential
well formation. Expressing the centrifugal term as $(angular momentum)^2/
(2ML^2)$ shows clearly that {\it only} the Thirring term contributes
to the correction in the
angular momentum of the "bound state". The physical reason is
the following. From (\ref{eqlfb}), notice that $B_\mu/g$ is identified as the
fermion current \cite{s}. The {\it vector nature} of the interaction leads
to the derivative-term upon integration, which subsequently changes the
angular momentum. Estimates of effective mass of the state can be obtained
from harmonic oscillator excitations around the well minimum.
\vskip .5cm
Acknowledgment: I am grateful to Professor S. Dutta Gupta, Director,
S. N. Bose National Centre for Basic Sciences, Calcutta, for allowing
me to use the Institution facilities.

\end{document}